\documentclass[%
 reprint,
superscriptaddress,
 amsmath,amssymb,
 aps,
]{revtex4-1}

\usepackage{graphicx}% Include figure files
\usepackage{dcolumn}% Align table columns on decimal point
\usepackage{bm}% bold math
\newcommand{\calB}{{\cal B}}
\newcommand{\calL}{{\cal L}}

\newcommand{\half}{{\frac{1}{2}}}

\begin{document}

\preprint{APS/123-QED}

\title{Anomalous Scaling of Stochastic Processes and the Moses Effect}% Force line breaks with \\

\author{Lijian Chen}
\affiliation{Department of Physics, University of Houston, Houston, TX 77204, USA}
\author{Kevin E. Bassler}
 \email{Email address: bassler@uh.edu}
\affiliation{Department of Physics, University of Houston, Houston, TX 77204, USA}
\affiliation{Department of Mathematics, University of Houston, Houston, TX 77204, USA}
\affiliation{Texas Center for Superconductivity, University of Houston, 77204, USA}
\author{Joseph L. McCauley}
\affiliation{Department of Physics, University of Houston, Houston, TX 77204, USA}
\author{Gemunu H. Gunaratne}
 \email{Email address: gemunu@uh.edu}
\affiliation{Department of Physics, University of Houston, Houston, TX 77204, USA}

\date{\today}% It is always \today, today,

\begin{abstract}
The state of a stochastic process evolving over a time $t$ is typically assumed to lie on a normal distribution whose width scales like $t^{\half}$. However, processes where the probability distribution is not normal and the scaling exponent differs from $\half$ are known. The search for possible origins of such ``anomalous" scaling and approaches to quantify them are the motivations for the work reported here. In processes with \emph{stationary increments}, where the stochastic process is time-independent, auto-correlations between increments and infinite variance of increments can cause anomalous scaling. These sources have been referred to as the \emph{Joseph effect} the \emph{Noah effect}, respectively. 
If the increments are \emph{non-stationary}, then scaling of increments with $t$ can also lead to anomalous scaling, a mechanism we refer to as the \emph{Moses effect}. 
Scaling exponents quantifying the three effects are defined and related to the 
Hurst exponent that characterizes the overall scaling of the stochastic process. 
Methods of time series analysis that enable accurate independent measurement 
of each exponent are presented.
Simple stochastic processes are used to illustrate each effect.
Intraday financial time series data is analyzed, revealing that its anomalous scaling 
is due only to the Moses effect. 
In the context of financial market data, we reiterate that the Joseph exponent, not the Hurst exponent, is the appropriate measure to test the efficient market hypothesis.  

\begin{description}
\item[PACS numbers]
02.50.Fz, 02.50.Ey, 89.65.Gh, 89.75.Da
\end{description}
\end{abstract}

\pacs{Valid PACS appear here}% PACS, the Physics and Astronomy
\maketitle

%\tableofcontents
\section{\label{sec:level1}Introduction}
A stochastic process is a sequence of random variables indexed 
either continuously or discretely, through a parameter often interpreted as time.
Stochastic processes have been used to model a range of phenomena from 
stock prices~\cite {Black1973, Mandelbrot1963,  mantegna1995scaling} to precipitation levels~\cite{le1961stochastic,richardson1981stochastic, wheater2005spatial} and animal locomotion \cite{viswanathan1999optimizing, sims2008scaling, soibam2014exploratory}. 
A standard example is Brownian motion, which is described by the Wiener process $\calB_t$. 
It has Gaussian distributed increments 
$\calB_{t+\tau}- \calB_t$ that are uncorrelated and 
independent of $t$.
The probability distribution of $\calB_t$ is also Gaussian, but with a width that grows as $t^{1/2}$.
Thus, $\calB_t$, also referred to as ``normal" diffusive motion~\cite{mandelbrot2002gaussian}, is said to \emph{scale} as $t^{\half}$.
More generally, processes are found that scale as $t^{H}$, 
where $H$ is referred to as the \emph{self-affine exponent} 
or \emph{Hurst exponent}~\cite{embrechts2002}. 

Following experimental observations including in biological systems~\cite{Iannaccone1995}, financial markets~\cite{mantegna1995scaling}, and turbulence~\cite{castaing1989scaling}, it is of considerable interest to understand the nature of stochastic processes that scale 
anomalously. 
For example, $H \neq 1/2$ has been associated with the failure of the efficient market hypothesis (EMH)~\cite{malkiel1970efficient, malkiel2003efficient}, namely that asset prices do not fully reflect all pertinent information  on the market~\cite{alvarez2008time,eom2008hurst,wang2010wti}.

For processes whose increments lie on stationary, time-independent probability distributions, Mandelbrot identified two
causes of anomalous scaling, 
which he referred to as the \emph{Noah Effect} and the \emph{Joseph Effect},
and furthermore defined scaling exponents to characterize them~\cite{mandelbrot1968noah, mandelbrot2002gaussian,Eliazar2013}.
The Noah effect represents the occurance of large
increments with anomalously large frequency resulting in a probability distribution with infinite variance. 
It is quantified by the \emph{latent exponent} $L$; 
increment distributions with $L > 1/2$ have ``fat tails" and exhibit anomalous scaling.
The Joseph effect occurs when increments are correlated 
and is quantified by the \emph{Joseph exponent} $J$. 
When $J \neq 1/2$ increment correlations can result in anomalous scaling.
Mandelbrot's nomenclature are biblical references: Noah built an ark
to save mankind and other creatures from the great flood~\cite{bible:Noah}, an occurrence of an anomalously large event, 
and Joseph, interpreting a dream of Pharaoh's, counceled him concerning what
he predicted would be a correlated sequence of years of abundance, 
followed by years of famine~\cite{bible:Joseph}.

In this paper, we extend the characterization of scaling of stochastic processes to include those 
with non-stationary increments. 
Such processes can model the intraday prices of financial markets~\cite{bassler2007nonstationary, muller1990statistical,dacorogna1993geographical},
daily precipitation levels~\cite{kantelhardt2006long, kantelhardt2003multifractality, brunsell2010multiscale, rehman2009study, richardson1981stochastic}, 
the abundance of solar flares~\cite{lepreti2000persistence, oliver1998there, rehman2009study, richardson1981stochastic},
and temperature fluctuations in turbulence~\cite{castaing1989scaling, rehman2009study, richardson1981stochastic}.
An additional mechanism that can produce anomalous scaling is 
identified. It is referred to as the \emph{Moses effect} and characterized with the \emph{Moses exponent} $M$,
which quantifies the growth of the increment distribution.
The nomenclature continues Mandlebrot's tradition: 
Moses led the Israelites after their Exodus from Egypt
as they wandered through the wilderness having no stationary 
settlements~\cite{bible:Moses}. 

The many abbreviations used in the text are summarized in Table~\ref{tab:table1}. 
\begin{table*}
\caption{\label{tab:table1} A list of abbreviations used in the text and their descriptions}
\begin{ruledtabular}
\begin{tabular}{|l|l|l|}
Abbreviations &Process or Index & Description \\ \hline
SIP & stationary increment process & increments: identical  \\
NIP & nonstationary increment process & increments: different  \\
CLT & central limit theorem &  \\
R/S & re-scaled range statistics & Eqn.~\eqref{rs_def}  \\
\hline
BM & Brownian motion & increments: independent, Gaussian distributed \\
LM & L\'evy motion & increments: independent, L\'evy distributed \\
FBM & fractional Brownian motion & increments: identical, correlated, Gaussian distributed \\
FLM & fractional L\'evy motion & increments: identical, correlated, L\'evy distributed \\
SBM & scaled Brownian motion & increments: scaled, independent, Gaussian distributed \\
SLM & scaled L\'evy motion & increments: scaled, independent, L\'evy distributed \\
SFBM & scaled fractional Brownian motion & increments: scaled, correlated, Gaussian distributed \\
SFLM & scaled fractional L\'evy motion & increments: scaled, correlated, L\'evy distributed \\
VDP & variable diffusion processes & Eqns.~\eqref{scaling_VDP} and ~\eqref{scaling_Diff} \\
\hline
DIA & Dow Jones Industrial Average & \\
SPY & S\&P500 Index & \\
QQQ & PowerShare NASDAQ-100 Index & \\
\end{tabular}
\end{ruledtabular}
\end{table*}

We will 
generalize the definitions of $H$, $L$, and $J$ and show that the sum of increments 
scale as $t^H$ with $H=L+J+M-1$.
For normal diffusive processes with stationary increments 
$L=J=M=1/2$ and thus $H=1/2$.
Anomalous scaling can occur when any of the exponents $L$, $J$, or $M$ differ from $1/2$.
We argue the even for processes with a Moses effect, 
as we have previously discussed~\cite{EMH1},
the EMH should be validated by measuring the Joseph exponent $J$, 
not the Hurst exponent $H$,
since the EMH relates to the absence of correlations in market returns. 
A measurement of $J \neq 1/2$, indicating the
presence of the Joseph effect, would violate the EMH,
but anomalous scaling resulting from a combination of 
Noah or Moses effects can still be
consistent with the EMH.

The principal result of the work is the decomposition of the overall scaling $H$ of the probability distributions into Joseph, Noah, and Moses effects via $H=J+L+M-1$.  Accurate numerical methods for establishing the four scaling exponents independently will 
be presented.
They account for finite-time corrections to the scaling behavior, and will be applied to
standard and scaled versions of example stochastic processes
to highlight the roles played by $H$, $L$, $J$, and $M$. 
Finally, exponents for a model of intraday trading in financial markets, 
variable diffusion processes~\cite{gunaratne2005variable, basslermarkov, mccauley2007hurst, mccauley2008martingales1, alejandro2006theory, bassler2007nonstationary, mccauley2009dynamics, Seemann2012, Hua2015}, 
will be computed and compared to the results with an empirical analysis of financial market data.   

The paper is organized as follows.
Definitions of the scaling exponents and methods to quantify them are given in Section 2.
A collection of example stochastic processes that will be studied and 
the numerical methods to simulate them are given in Section 3. 
Section 4 presents the analysis of scaling in the various processes, 
demonstrating how the different effects can combine to yield an overall anomalous scaling.
The methods used to accurately  measure scaling indices are also presented in this section. 
In section 5, empirical financial market data is analyzed and compared to that of
variable diffusion processes. The results are discussed in Section 6.

\section{Scaling in Stochastic Processes}
Consider a one-dimensional stochastic process $\{X_t;t\in \mathcal{T}\}$ where 
the set $\mathcal{T}$ can either be a subset of the real numbers or 
a subset of the integers.
If the increments of the process
\begin{equation}
\delta_t(\tau)
=
X_{t+\tau} - X_{t}
\label{deltadef}
\end{equation}
have a probability distribution that is independent of $t$,
then 
the stochastic process is said to be a stationary increment process (SIP).
If instead, the probability distribution of $\delta_t(\tau)$ depends on $t$, 
the process is referred to as a non-stationary increment process (NIP). 
Statistical analyses of SIPs can be preformed on a single time series of data,
as the time independence of the increments permits a statistical ensemble to be
constructed by time translations of the starting point.
Statistical analyses of NIPs, on the other hand, generally can not be performed
on a single time series, requiring instead an ensemble of time series. 
If, however, a NIP 
%repeats itself, either periodically or otherwise, so that it 
begins anew at certain times, such as after a triggering event, then corresponding time translations may be used to construct a statistical ensemble from a single time-series~\cite{mccauley2008time, bassler2007nonstationary, mccauley2008martingales2}. 
Although, this perhaps can only be done if the time between renewals has a finite average~\cite{sibani2013}, since, more generally, weak ergodicity breaking~\cite{bouchaud1992,bel2005} 
may prevent time averages being equated with ensemble averages
in diffusive processes that scale anomalously. 

The generalization of time series analyses to include NIPs necessitates generalizing the
definitions of indices used to characterize scaling in SIPs.
Consider an ensemble of realizations of a stochastic process
$\mathcal{X}=\{X^{(p)}_t, p=1,2,....\}$, each of which
starts at the origin, \emph{i.e.,} $X^{(p)}_0=0$, and
has increments $\delta_t^{(p)}(\tau)$.
The increments are random variables with probability distributions that can
depend on $t$, but are the same for all realizations $p$.
These realizations can either have continuous or discrete time, but for the purposes of
analyzing the scaling of the process
assume that they are sampled at regular intervals of time $\tau$,
which can be taken to be unity. 
The sampled times are then $t=1,2,3,\ldots$.
%Also assuming that each realization begins at the origin,
%\emph{i.e.}, $X^{(p)}_0=0$ for all $p$,
%at integer values of $t$, 
%each process is the sum of its increments
Then
\begin{equation}
X_t = \sum\limits_{s=0}^{t-1}\delta_s ,
\label{X}
\end{equation}
where here and in what follows the $p$ superscript is suppressed for
simplicity and $\delta_t=\delta_t(1)$. 
Define also the following random variables:
the the sum of the absolute values of increments
\begin{equation}
Y_t = \sum\limits_{s=0}^{t-1}|\delta_s| ,
\label{A}
\end{equation}
and the sum of increment squares
\begin{equation}
Z_t = \sum\limits_{s=0}^{t-1} \delta^{2}_s.
\label{Z}
\end{equation}
Probability distributions of these variables and of $X_t$
over the ensemble $\mathcal{X}$ will be used to characterize and
quantify the scaling of NIPs.  
The definitions of the scaling exponents that follow, which  
involve ensemble averages, \emph{i.e.}, over $p$, 
become equivalent to standard definitions for processes with stationary increments.
%Averages over the ensemble $\mathcal{X}$ estimate stochastic averages over the corresponding NIP.
%It is important though to realize that these two types of averages are not equivalent for NIPs.

A stochastic processes $X$ is \emph{self-similar} if, for any $a>0$, 
there exists an exponent $H \geq 0$ such that
\begin{equation} X_{at}  \stackrel{d}{=}  a^{H}X_t 
\label{selfsimilar}
\end{equation}
where ``$\stackrel{d}{=}$" represents equality ``in distribution;"
$H$ quantifies the scaling of the overall process.
Note that, in general, only the one-point probability distribution $W(X_t)$  
scales in self-similar processes; higher-order, multi-point distributions do not necessarily scale.
Operationally, scaling of a suitable  ``width" of the probability distribution of $X_t$ can be used 
to estimate $H$ 
\begin{equation}
w[X_t] \sim t^{H}.
\label{DefH}
\end{equation}
In this paper, we use the difference of the 75th quantile and the 25th quantile
of the ensemble probability distribution as the measure of this width. 
For SIPs this definition of $H$ becomes equivalent to Mandelbrot's \cite{embrechts2002,mandelbrot2002gaussian,Eliazar2013}
and can be estimated by a variety of means, 
including de-trended fluctuation analysis \cite{peng1994mosaic}. Using the quantiles circumvents difficulties in cases where moments of the probability distribution diverge.
%(The DFA measures H. But, the DFA use mean, which is not robust when increments are non-Gaussian, if we modified DFA by using median instead, it will be robust.--Lijian)

%Often, discrete stochastic processes will only be assymtotically self-similar. 
%In this case, each increment $X_s$ in Eqs.\ref{Y}-\ref{A} must themselves be the sum of
%an assytomtically large number of increments $x$, i.e.
%\begin{equation}
%X_s = \sum_{i=1}^l x_i \nonumber
%\end{equation}
%where $l \gg 1$.

Since $X_t$ is the sum of increments [Eq.~\eqref{X}], 
it follows from 
the central limit theorem (CLT) that 
if the increment probability distributions are 
\begin{itemize}
\item[(a)] uncorrelated,
\item[(b)] have finite variance, and
\item[(c)] identical, independent of time,
\end{itemize}
then
the process will scale ``normally," with
$H=\frac{1}{2}$.
Anomalous scaling in stochastic processes ($H\neq \frac{1}{2}$) may originate from the failure of one or more of these conditions.
Joseph, Noah and Moses effects are associated with the failures of conditions (a), (b), and (c) respectively, which 
are analyzed in following subsections.

\subsection{Joseph effect}
The \emph{Joseph effect} is associated with the failure of condition (a) 
and can be quantified through 
a variant of rescaled range statistics (R/S)~\cite{mandelbrot1969robustness, mandelbrot1968noah, avram2000robustness, mandelbrot2002gaussian, hurst1951long}. 
Estimate the range $R_t$ and standard deviation $S_t$ 
of a stochastic process as
\begin{equation}
\begin{aligned}
R_t & =\max\limits_{1 \leq s \leq t} \left[X_s - \frac{s}{t}X_t \right] 
- \min\limits_{1 \leq s \leq t} \left[X_s - \frac{s}{t}X_t \right]\\
S^{2}_t & = \frac{1}{t} Z_t - \left[\frac{1}{t}X_t\right]^2 .
\label{rs_def}
\end{aligned}
\end{equation}
Then the
ensemble averaged 
ratio of $R_t$ and $S_t$
scales as
\begin{equation}
E\left[R_t/S_t\right] \sim t^{J},
\label{DefJ}
\end{equation} 
where $J$ is the \emph{Joseph exponent}.
Negatively dependent processes have $J\in (0,1/2)$, positively dependent processes have $J\in (1/2,1)$, and independent processes have $J = 1/2$.

%The Equations are defined in the range $\{Y(s), 1\leq s\leq t \}$ which fixing the starting point at $s=1$, they applied to both SIPs and NIPs. For SIPs, the average can be taken over different starting points instead \cite{mandelbrot1969robustness}. 

\subsection{Noah effect}
The \emph{Noah effect} refers to the failure of condition (b), and is
quantified by the \emph{latent exponent} $L$ \cite{mandelbrot1968noah, mandelbrot2002gaussian, Eliazar2013}. 
Suppose the tails of the increment distributions decay as
\begin{equation}
\lim_{x \rightarrow \infty} Pr(|\delta_t|>x) \sim x^{-\gamma},
\end{equation}
where $\gamma\in (0,\infty)$. Then $L(t) := max(\frac{1}{2}, \frac{1}{\gamma})$. In this paper, we only consider processes for which $L$ is independent of time. 
Note that CLT condition (b) fails when $L>\frac{1}{2}$, 
because the variance of $\delta_t$ is then infinite. 
If the increment distribution is Gaussian, log-normal or is 
any distribution with $\gamma \geq 2$, then $L=\frac{1}{2}$. 
If instead the increment distribution has fat tails with $\gamma<2$, then 
$L = \frac{1}{\gamma}$. 
We further limit our considerations to processes with $L < 1$,
as otherwise the increment distributions 
have infinite mean \cite{samoradnitsky1994stable}. 

For time series analyses, 
a more convenient and stable way to estimate the latent exponent is from the 
scaling of ensemble probability distribution of the sum of increment squares,
which can be estimated by the scaling of the median of the probability distribution of $Z_t$
\begin{equation}
    m\left[Z_t\right] \sim t^{2L+2M-1}
\label{DefLM}
\end{equation}
where 
$M$ is \emph{Moses exponent} introduced below. 
A proof for Eqn.~\eqref{DefLM} is given in the Supplemental Material at [URL will be inserted by publisher]. 

\subsection{Moses effect}
We define the failure of condition (c) as the \emph{Moses effect}. 
It occurs when the increment distribution is time dependent. 
We consider processes 
with increment distributions whose mean absolute deviation scale as
\begin{equation}
%E\left[\delta^2_t - E^2[\delta_t]\right] \sim t^{2M-1},
E\left[\left|\delta_t - E(\delta_t)\right|\right] \sim t^{M-\frac{1}{2}}
\end{equation}
%(I think variance is not the best here, since variance is not exists for non-Gaussian increments, the mean absolute value is better, since it still exists for non-Gaussian with 1< alpha<2.---Lijian)
where, as before, $E$ is the ensemble average.
For SIPs, $M=\frac{1}{2}$, whereas for NIPs, $M\neq\frac{1}{2}$.

In time series analyses,
a convenient and more robust way to estimate
the Moses exponent is from the scaling of the
ensemble probability distribution of the sum of the absolute value of increments,
which can be estimated by the scaling of the median of the probability distribution of $Y_t$
\begin{equation}
    m\left[Y_t\right] \sim t^{M + \frac{1}{2}}.
    \label{DefM}
\end{equation}
A proof for Eq.~\eqref{DefM} and a discussion of the effects 
on changes in the measurement frequency are given in the Supplemental Material at [URL will be inserted by publisher].
Typically, we find that varying the increment interval, $\tau$ in Eqn.~\ref{deltadef},
does not affect the leading scaling behavior of $Y_t$.
% m[Y_t] or Y_{(0.5)}(t) ???

The anomalous scaling of NIPs can arise due to combination of all three effects listed above. 
Equations~\eqref{DefH}, \eqref{DefJ}, \eqref{DefLM}, and \eqref{DefM}
provide independent estimates of the four exponents.
However, they are related through
\begin{equation}
H = J+L+M-1.
\label{equality}
\end{equation}
This scaling relation provides a useful independent check of the
estimates of the four exponents.

%Note that \emph{Hurst exponent} $H$ has been traditionally used to characterize scaling in SIPs and is defined as $H = J + L - 1/2$ \cite{mandelbrot2002gaussian}. The scaling of NIPs which can arise due to combination of all three effects listed above. It's given by $H = J+L+M-1$ as we will demonstrated through examples given in the next section.

\section{Examples of Self similar processes}

In this section several model stochastic processes are introduced that will be used to illustrate the relationship~\eqref{equality}. However, we emphasize that the relationship is expected to be valid for other stochastic processes as well. 

\subsection{Processes with Gaussian Increments}

The classic example is \emph{Brownian motion} (BM) which consist of a sequence of identical, independent Gaussian increments~\cite{Feller:1957}. BM can be generalized by including (a) correlations between increments, and (b) time-dependent increments. One way to include long-term correlations is through \emph{fractional Brownian motion} (FBM),  denoted $\calB^{(J, \frac{1}{2})}_t$, and defined below. These ``correlations" can be characterized by an index $J$ which, as shown below, is the Joseph exponent. As for time-dependent increments, we limit consideration to processes whose increments scale in time. As shown below, the growth can be characterized by the Moses exponent $M$. A stochastic process consisting of correlated and scaled Gaussian increments is denoted $\calB^{(J,M)}_t$. In this notation, Brownian motion $\calB_t$ is denoted $\calB^{(\frac{1}{2}, \frac{1}{2})}_t$.

Fractional Brownian motions $\calB^{(J, \frac{1}{2})}_t$ that scale in time have the form  
\begin{eqnarray}
\calB^{(J, \half)}_t \propto & \int _{-\infty} ^{0} \left[ (t-s)^{J-\half} -(-s)^{J-\half} \right] d\calB_s \nonumber\\
    &+ \int_0^t (t-s)^{J-\half} d\calB_s,
\end{eqnarray}
where the the proportionality constant is $1/{\Gamma(J+\half)}$ and 
$d\calB_s \equiv \calB_{s+ds} - \calB_s = \calB_{ds}$. The Gaussian increments of FBM are correlated~\cite{mandelbrot1975limit, avram1986some}, and the scaling exponents are derived as follows:
\begin{itemize}
\item[(i)] Using Donsker's theorem~\cite{donsker1951invariance} and continuous mapping therom \cite{billingsley2013convergence, whitt2002stochastic}, Avram and et al. proved that the index $J$ is the \emph{Joseph exponent} \cite{mandelbrot1975limit, avram1986some, avram2000robustness}. 
\item[(ii)] Since individual increments are Gaussian-distributed, $L=1/2$. 
\item[(iii)] Since FBM is a SIP, $M=\frac{1}{2}$. 
\item[(iv)] To derive $H$, let $v = s/t$, then $ds = t dv$, $d\calB_s \stackrel{d}{=} \calB_{ds} \stackrel{d}{=} \calB_{tdv} \stackrel{d}{=} t^{\frac{1}{2}}d\calB_{v}  $. Thus, $\calB^{(J, \frac{1}{2})}_t = \int _{-\infty} ^{t}(t-s)^{J-\frac{1}{2}}d\calB_s \stackrel{d}{=} \int _{-\infty} ^{1} (1-v)^{J-\frac{1}{2}} t^{J-\frac{1}{2}} t^{\frac{1}{2}} d\calB_v \stackrel{d}{=} t^{J} \int _{-\infty} ^{1} (1-v)^{J-\frac{1}{2}} d\calB_v  \stackrel{d}{=}  t^{J}\calB^{(J, \frac{1}{2})}_1$. From definition ~\eqref{selfsimilar} $H = J$. 
\end{itemize}
Observe that $H = J+L+M-1$.

A scaled FBM is an NIP, except when $M=\frac{1}{2}$, and is defined as
\begin{equation}
\calB^{(J,M)}_t = \int_0^t s^{M-\frac{1}{2}} d\calB^{(J, \frac{1}{2})}_s,
\end{equation}
where $d\calB^{(J, \frac{1}{2})}_s = d\calB^{(J, \frac{1}{2})}_{s+ds} - d\calB^{(J, \frac{1}{2})}_s$. Since increments $d \calB^{(J,M)}_t = t^{M-\frac{1}{2}} d\calB^{(J, \frac{1}{2})}_t$ and FBM is a SIP, $M$ is the Moses exponent of $\calB^{(J,M)}_t$. Furthermore, since $\calB^{(J,M)}_t = t^{M-\frac{1}{2}} \calB^{(J, \frac{1}{2})}_t = t^{J+M-\frac{1}{2}} \calB^{(J, \frac{1}{2})}_1$ [see (iv) above], the self-similarity exponent is $H=J+M-\frac{1}{2}$. Consequently, $H = J+L+M-1$.

\emph{Generation of FBM and scaled FBM:} There are several methods proposed to generate FBM, three of them, the Hosking method~\cite{brockwell2013time}, the Cholesky method~\cite{dieker2004simulation, bardet2003generators}, and the Davies-Harte method~\cite{davies1987tests, dietrich1997fast, wood1994simulation}, are exact. We used the Davies-Harte method, which  requires $O(N\log\ N)$ operations, due to its computational efficiency. The algorithm is predicated on computing the square root of the covariance matrix using the circulant matrix and a fast Fourier transform instead of much slower lower-upper triangular decomposition~\cite{dieker2004simulation, bardet2003generators}.

\subsection{Processes with L\'evy Increments}

Independent and identically distributed increments in the classic L\'evy motions can be characterized by a single index $L$~\cite{samoradnitsky1994stable}, which, as shown below, turns out to be the latent index. As before we can generalize the underlying process by including correlations and time-scaling of increments. The resulting process will be denoted $\calL^{(J,L,M)}_t$. 

Increments of the L\'evy motion $\calL^{(\half, L,  \half)}_{t}$ are stochastic variates from a probability distribution whose characteristic function is~\cite{gnedenko1954limit,uchaikin1999chance, nolan2015} 
\begin{equation}
E[\exp(i\theta \delta)] = \exp \left[-|\theta|^{(1/L)} \right].
\end{equation}
The scaling exponents for L\'evy motions are evaluated as follows:
\begin{itemize}
\item[(i)] It has been shown that ${\calL^{(\half, L,  \half)}_{t}}/{t^{L}} \stackrel{d}{=} \calL^{(\half, L,  \half)}_1$, where $t \geq 0$ \cite{embrechts2002, samoradnitsky1994stable}. Consequently, $R(t) \sim t^{L} $. For $L > \half$, increments of LM have the property that $Pr(|\delta_{t}|>x) \sim x^{-1/L}$, which implies that $Pr(\delta_{t}^{2}> x^{2} = y) \sim x^{-1/L} = y^{-1/2L}$. Applying CLT for processes with infinite variance~\cite{uchaikin1999chance, gnedenko1954limit}, we obtain that $Z(t) \sim t^{2L}$, $S(t) = \sqrt{Z(t)/t} \sim t^{L-1/2}$. Thus, $E_{e}\left[R(t)/S(t)\right] \sim t^{1/2}$, and $J=\frac{1}{2}$~\cite{mandelbrot2002gaussian}.
\item[(ii)] Increments satisfy  $Pr(|\delta_{t}|>x) \sim x^{-1/L}$ implying that $L$ is the latent exponent. 
\item[(iii)] For $\delta_{t}$ with $L < 1$, the first order moment is bounded; since LM is a SIP, we have $E[|\delta_{t} - E(\delta_{t})|] \sim t^{0}$, therefore $M=\frac{1}{2}$. 
\item[(iv)] $H = L$ ~\cite{embrechts2002, samoradnitsky1994stable}.
\end{itemize}
Note that $H = L + J + M -1$.

As with the Brownian case, correlations between the variates can be induced using the \emph{Fractional L\'evy motion} (FLM) defined as
\begin{eqnarray}
\calL^{(J, L, \half)}_t \propto&  \int _{-\infty}^{0} \left[ (t-s)^{J-\half} -(-s)^{J-\half} \right] d\calL^{(\half, L,  \half)}_s \nonumber\\
     &+ \int_0^t (t-s)^{J-\half} d\calL^{(\half, L,  \half)}_s,
\label{FLM_Def}
\end{eqnarray}
where $d\calL^{(\half, L,  \half)}_s = \calL^{(\half, L,  \half)}_{s+ds} - \calL^{(\half, L,  \half)}_s = \calL^{(\half, L,  \half)}_{ds}$ and the proportionality constant is $ 1/{\Gamma(J+\half)} $. 
Only the exponents $J$ and $H$ of a FLM differ from those of the corresponding L\'evy motion. The Joseph exponent cannot be defined using R/S statistics when $J < \frac{1}{2}$, since the process is nowhere bounded in that case~\cite{embrechts2002,samoradnitsky1994stable,Eliazar2013}.  Thus, we restrict consideration to $J \geq \frac{1}{2}$. Avram et al. proved that the index $J$ here is Joseph exponent~\cite{avram1986some, avram2000robustness}. 

In order to estimate H, setting $s=av$, Eqn.~\eqref{FLM_Def} can be expressed as 
\begin{eqnarray}
\calL^{(J, L, \half)}_{at} \stackrel{d} \propto& \int _{-\infty}^{0} \left[ (at-av)^{J-\half} -(-av)^{J-\half} \right] a^L d\calL^{(\half, L,  \half)}_v \nonumber\\
     &+ \int_0^t (at-av)^{J-\half} a^L d\calL^{(\half, L,  \half)}_v,\nonumber
\end{eqnarray}
where the proportionality constant is the same as before. It follows that $\calL^{(J, L, \half)}_{at} \stackrel{d} \propto a^{J+L-\half} \calL^{(J, L, \half)}_{t}$, and hence that $H=J+L-\half$. 

Finally, time-dependent increments can be generated through scaled FLM is defined via
\begin{equation}
    \calL^{(J,L,M)}_t = \int_{0}^{t} s^{M-\frac{1}{2}} d\calL^{(J,L,\half)}_s, \nonumber
\end{equation}
where $d\calL^{(J, L,  \half)}_s = \calL^{(J, L,  \half)}_{s+ds} - \calL^{(J, L,  \half)}_s$.
Only the values of $M$ and $H$ of a scaled FBM differ from the corresponding exponents of the associated FBM. Specifically,  using a calculation similar to that of $H$ for FBM, $\calL^{(J,L, M)}_{t} \stackrel{d}{=} t^{J+L + M - 1} \calL^{(J,L, M)}_{1}$, thus $H = J+L + M - 1$. 

\emph{Generation of L\'evy-stable random variables.} Let $\epsilon$ be a uniform random variate on $(-\frac{\pi}{2}, \frac{\pi}{2})$ and let a random variable $\Phi$ be exponential with mean 1. Assume $\epsilon$ and $\Phi$ to be independent. Then the random variable
\begin{equation}
X = \frac{\sin (\epsilon/L)} {(\cos \epsilon)^L} \left( \frac{\cos((L-1)\epsilon / L)} {\Phi} \right)^{(L-1)/L} \nonumber
\end{equation}
is known to be distributed $\calL(\half, L,  \half)$~ \cite{samoradnitsky1994stable, dumouchel1971stable, chambers1976method}.

The Davies-Harte method cannot be used to generate FLM variates since the associated correlation function does not exist. We used an approach introduced by Stoev  and  Wu~\cite{stoev2004simulation, wu2004simulating} that takes advantage of the circulant matrix and the fast Fourier transform to  generate  FLM. It requires $O(N\log N)$ operations. However, the algorithm can only simulate FLM approximately.

\subsection{Variable Diffusion Process}

Processes with Gaussian or L\'evy increments, discussed above, have one common characteristic, that the increments are from a stable process and independent of the stochastic  variable $X_t$. In this subsection we introduce a set of diffusive processes whose increments depend on $X_t$ as well.  \emph{Variable diffusion processes} (VDPs) was introduced as a model for intraday variations in financial markets~\cite{bassler2007nonstationary, Seemann2012, Hua2015}. Here $\{X_t, t\geq 0\}$ satisfies the stochastic differential equation: $dX_t = \sqrt{D[X_t, t]} d\calB_{t}$, where $D(X_t, t)$ is the diffusion coefficient.  If the probability distribution function $W(X_t)$ at time $t$ is self-similar (as given by Eqn.~\eqref{selfsimilar}), 
\begin{equation}
    W(X_t) = t^{-H} {\cal F}(u),
    \label{scaling_VDP}
\end{equation}
where the \emph{scaling variable} is $u = {X_t}/{t^{H}}$. Variable diffusion processes exhibit many \emph{stylized facts} (\emph{i.e.,} common statistical features) reported in financial markets~\cite{gunaratne2005variable}.

For variable diffusion processes with finite variance, Eqn.~\eqref{scaling_VDP} shows that $E[X_t] \sim t^H$ and $E[X_t^2] \sim t^{2H}$, and hence that  $Var[X_t] =t^{2H}Var[X_1]$. The self-similarity of the probability distribution implies further that the diffusion coefficient scales as~\cite{gunaratne2005variable}
\begin{equation}
    D(X_t, t) = t^{2H-1} {\cal D}(u).
    \label{scaling_Diff}
\end{equation} 
The probability distribution $W(X_t)$ satisfies the Fokker-Planck equation:
\begin{equation}
\frac{\partial}{\partial (t)} W(X_t) = \frac{1}{2} \frac{\partial^{2}}{\partial Y^{2}}\left[D(X_t, t) W(X_t)\right].
\end{equation}
Using Eqns.~\eqref{scaling_VDP} and ~\eqref{scaling_Diff}, we obtain $2H(uf(u))' +(D(u)f(u))'' = 0$, 
whose solution is 
\begin{equation}
{\cal F}(u) = \frac{C}{{\cal D}(u)}\exp \left(-2H \int \frac{u du}{{\cal D}(u)} \right)
\end{equation}
As an example, if ${\cal D}(u)$ is constant $D_{0}$, ${\cal F}(u) = C_{0} \exp \left(-\frac{1}{2D_{0}}u^{2} \right)$. If ${\cal D}(u) = D_{0}(1+\epsilon |u|)$ and $D_{0}=\frac{2H}{\epsilon ^{2}}$, where $\epsilon$ is a constant, then ${\cal F}(u) = \frac{\epsilon}{2} \exp (-\epsilon |u|)$, the bi-exponential distribution. The corresponding variable diffusion process $\{X_{H}(t), t\geq 0\}$ is given by
\begin{equation}
X_t = \int_{0}^{t} s^{H-\frac{1}{2}} \sqrt{\frac{2H}{\epsilon^{2}} \left(1 + \epsilon \left| \frac{X_s}{s^{H}} \right| \right)} d\calB_s. \nonumber
\end{equation}
The associated exponents are:
\begin{itemize}
\item[(i)] Since VDP is a Markov process $J=\frac{1}{2}$.
\item[(ii)] For VDP with finite variance  $L=\frac{1}{2}$.
\item[(iii)] $dX_t= \sqrt{D(X_t, t)} d\calB_t = t^{H-\frac{1}{2}}\sqrt{{\cal D}(u)} d\calB_t$, thus $M=H$.
\item[(iv)] Since $W(X_t) = W(t^{H}X_1)$, the Hurst exponent is $H$. 
\end{itemize}
Once again, $H=J+L+M-1$.

\section{Results From Simulations}
\subsection{Finite-Size Corrections}
\begin{figure*}
\includegraphics[width = 0.80\textwidth]{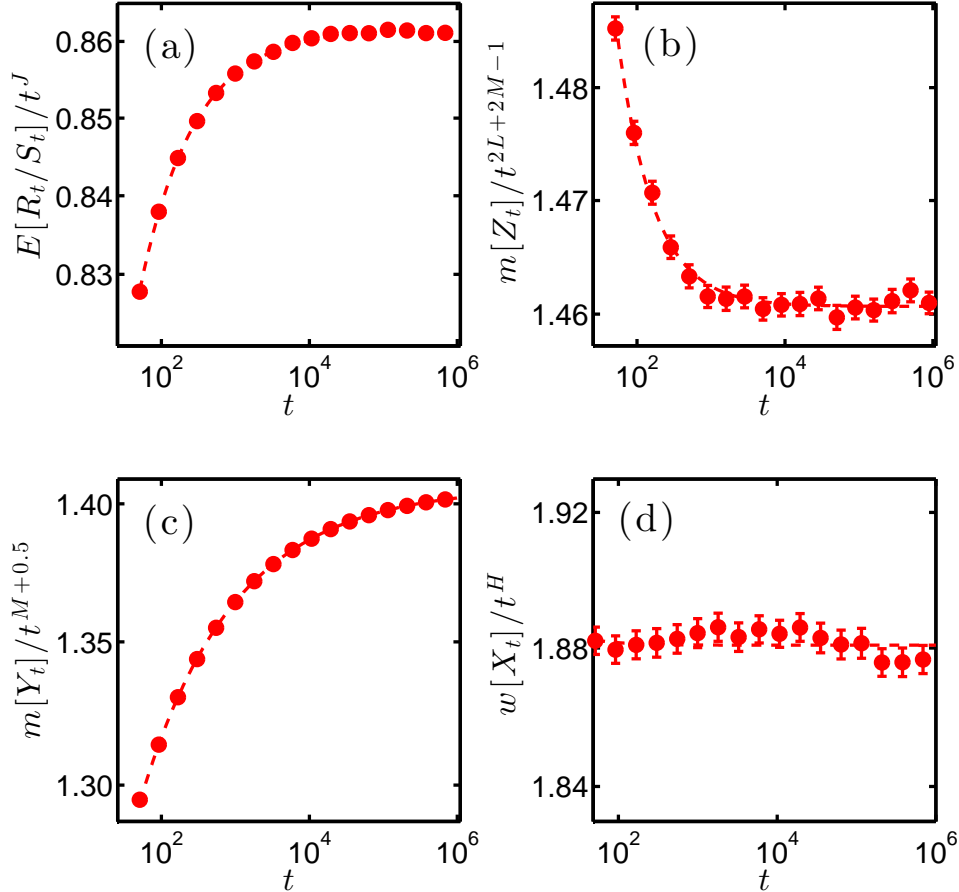}% Here is how to import EPS art

\caption{(Color online) Nonlinear fitting for SFLM ($J=0.6$, $\alpha=1.6666, M=0.6$) using the form $y(t)/t^{\Omega} = a + b t^{-c}$ for finite-time corrections. (a) $y(t) = E_{e}[R_t/S_t]$, $\Omega = J$. (b) $y(t) = m[Z_t]$, $\Omega = 2L+2M-1$. (c) $y(t)=m[Y_t]$, $\Omega = M+0.5$. (d) $y(t)=w[X_t]$, $\Omega = H$.}
\label{fig:SFLM}
\end{figure*}

Rescaled range statistics (R/S) analysis has been used extensively in studying persistence and long-term dependence in natural time series. The classical approach using the best linear relationship between $\log[E(R/S(t))]$ and $\log(t)$ yields a biased estimate unless $t$ is large~\cite{caccia1997analyzing,bisaglia1998comparison, taqqu1995estimators, hamed2007improved}. Corresponding approaches to measure $L$, $M$, and $H$ 
also suffer from analogous finite-time corrections. Ref.~\cite{hamed2007improved}  showed that the first order finite-time corrections take the form
\begin{equation}
    y(t)/t^{\Omega} = a + b t^{-c},
    \label{FTC0}
\end{equation}
where $a,b,c$ are constants. As an example, Figure~\ref{fig:SFLM} shows the nonlinear fit given by Eqn.~\eqref{FTC0} for a SFLM with parameters $J=0.6$, $\alpha=1.6666$, and $M=0.6$ for the (known) indices $J$, $2(L+M-\half)$, $M+\half$, and $H$. Here, the reciprocal of the variance has been used as the weight of a point. 

When exponents for a stochastic process are unknown (\emph{e.g.,} financial markets) we use the form
\begin{equation}
    y(t) = a t^{\Omega^{\prime}} + b t^{\Omega^{\prime}-c}
    \label{FTC}
\end{equation}
for finite-time corrections, and estimate $\Omega^{\prime}$ as well. For the remainder of the paper, we use this approach to estimate the exponents $H$, $J$, $L$, and $M$.

\subsection{Exponents for different processes}
For model stochastic processes, we use an ensemble of 100,000 stochastic realizations, each of length $t_{max}=1$ million. The lower cutoff is chosen to be $t_{min} = 50$. We select $t_{\#}=500$ points between $t_{min}$ and $t_{max}$; the i-th point $t(i)$ ($i \in [1,2, ... t_{\#}]$) is  
\begin{equation}
    t(i) =  round\left[t_{min} \left( \frac{t_{max}}{t_{min}} \right)^{\frac{i}{t_{\#}}} \right],
    \label{times}
\end{equation}
where $round[x]$ represents the integer nearest to $x$; these points are (approximately) uniformly distributed in log scale. $J$, $L+M$, $M$, $H$, estimated independently, are given in Table~\ref{tab:table2}. The reported standard errors are obtained through a bootstrap method. The relation $H = L+M+J-1$ is found to hold for each process. 

\begin{table*}
\caption{\label{tab:table2} Estimates of the exponents for several classes of self similar processes using the form $y(t) = a t^{\Omega^{\prime}} + b t^{\Omega^{\prime}-c}$ for finite-time corrections. Note that when $J < \half$, it cannot be evaluated using the R/S method~\cite{embrechts2002}.}
\begin{ruledtabular}
\begin{tabular}{|c|c|c|c|c|c|}
Processes  &$J$ &$L$  &$M$ &$J+L+M-1$ &$H$ \\ \hline
BM   &0.4997(5) &0.5000(1) &0.5000(1) &0.4997(5) &0.501(1) \\ 
SBM($M=0.3$)   &0.4996(5) &0.5000(1) &0.3000(1) &0.2996(5) &0.300(3) \\ 
SBM($M=0.4$)   &0.4999(5) &0.5000(1) &0.4000(1) &0.3999(5) &0.400(1) \\ 
SBM($M=0.6$)   &0.4995(4) &0.5000(1) &0.6000(1) &0.5995(5) &0.601(1) \\ 
SBM($M=0.7$)   &0.4993(4) &0.5000(1) &0.7000(1) &0.6993(5) &0.700(2) \\ 
LM($L=0.71$)   &0.4998(2) &0.7139(6) &0.5001(2) &0.7139(6) &0.714(2) \\ 
LM($L=0.58$)   &0.4996(3) &0.5883(2) &0.5000(1) &0.588(3) &0.588(3) \\ 
LM($L=0.53$)   &0.5004(4) &0.5266(2) &0.5000(1) &0.5270(5) &0.5256(9) \\  
SLM($L=0.53$, $M=0.3$)   &0.4998(3) &0.5268(3) &0.3000(1) &0.3266(6) &0.3254(4) \\ 
SLM($L=0.53$, $M=0.4$)   &0.4999(7) &0.5263(2) &0.4000(1) &0.4263(9) &0.426(2) \\ 
SLM($L=0.53$, $M=0.6$)   &0.4989(6) &0.5265(2) &0.6000(1) &0.6253(8) &0.626(2) \\ 
SLM($L=0.53$, $M=0.7$)   &0.5000(4) &0.5265(2) &0.7000(1) &0.7265(5) &0.726(1) \\
SLM($L=0.77$, $M=0.3$)   &0.5000(2) &0.769(5) &0.3004(4) &0.569(5) &0.569(1) \\ 
SLM($L=0.77$, $M=0.4$)   &0.4998(2) &0.768(2) &0.4009(5) &0.669(2) &0.669(2) \\ 
SLM($L=0.77$, $M=0.6$)   &0.5002(2) &0.769(5) &0.6006(5) &0.870(4) &0.869(2) \\ 
SLM($L=0.77$, $M=0.7$)   &0.4997(2) &0.768(4) &0.7006(5) &0.969(3) &0.970(1) \\
FBM($J=0.3$)   &0.2994(5) &0.5000(1) &0.5000(1) &0.2994(5) &0.300(1) \\ 
FBM($J=0.4$)   &0.4000(5) &0.5000(1) &0.5000(1) &0.3999(5) &0.400(1) \\ 
FBM($J=0.6$)   &0.6001(3) &0.5000(1) &0.5000(1) &0.6001(3) &0.598(2) \\ 
FBM($J=0.7$)   &0.6998(2) &0.5000(1) &0.5000(1) &0.6998(2) &0.700(1) \\ 
SFBM($J=0.3$, $M=0.3$)   &0.2988(7) &0.5000(1) &0.3000(1) &0.0988(8) &0.097(3) \\ 
SFBM($J=0.3$, $M=0.4$)   &0.3004(5) &0.5000(1) &0.4000(1) &0.2004(5) &0.201(2) \\ 
SFBM($J=0.3$, $M=0.6$)   &0.2997(5) &0.5000(1) &0.6000(1) &0.3998(6) &0.400(2) \\ 
SFBM($J=0.3$, $M=0.7$)   &0.2989(6) &0.5000(1) &0.7000(1) &0.4989(6) &0.500(1) \\ 
SFBM($J=0.4$, $M=0.3$)   &0.3990(6) &0.5000(1) &0.3000(1) &0.1990(6) &0.2007(5) \\ 
SFBM($J=0.4$, $M=0.4$)   &0.3991(6) &0.5000(1) &0.4000(1) &0.2991(6) &0.3006(4) \\ 
SFBM($J=0.4$, $M=0.6$)   &0.3993(4) &0.5000(1) &0.6000(1) &0.4993(4) &0.500(1) \\ 
SFBM($J=0.4$, $M=0.7$)   &0.3994(4) &0.5000(1) &0.7000(1) &0.59994(4) &0.6001(5) \\ 
SFBM($J=0.6$, $M=0.3$)   &0.5998(5) &0.5000(1) &0.3000(1) &0.3998(5) &0.4004(6) \\ 
SFBM($J=0.6$, $M=0.4$)   &0.5998(3) &0.5000(1) &0.4000(1) &0.4998(3) &0.5004(5) \\ 
SFBM($J=0.6$, $M=0.6$)   &0.5998(2) &0.5000(1) &0.6000(1) &0.6998(2) &0.700(1) \\ 
SFBM($J=0.6$, $M=0.7$)   &0.6000(3) &0.5000(1) &0.7000(1) &0.8000(3) &0.800(2) \\ 
SFBM($J=0.7$, $M=0.3$)   &0.6989(8) &0.5000(1) &0.3000(1) &0.4989(8) &0.498(2) \\ 
SFBM($J=0.7$, $M=0.4$)   &0.6999(3) &0.5000(1) &0.4000(1) &0.5999(3) &0.5987(7) \\ 
SFBM($J=0.7$, $M=0.6$)   &0.7000(2) &0.5000(1) &0.6000(1) &0.7999(2) &0.7997(4) \\ 
SFBM($J=0.7$, $M=0.7$)   &0.6999(1) &0.5000(1) &0.7000(1) &0.8999(2) &0.900(2) \\ 
FLM($J=0.4$, $L=0.60$)  &- &0.600(4) &0.4999(1) &- &0.4991(5) \\ 
FLM($J=0.6$, $L=0.60$)  &0.5996(3) &0.600(3) &0.4999(1) &0.699(3) &0.699(1) \\ 
FLM($J=0.4$, $L=0.56$)  &- &0.5560(3) &0.4999(1) &- &0.4564(7) \\ 
FLM($J=0.6$, $L=0.56$)  &0.5996(3) &0.5558(4) &0.5000(1) &0.6553(7) &0.6547(4) \\ 
SFLM($J=0.4$, $L=0.60$, $M=0.3$) &- &0.601(2) &0.2998(1) &- &0.300(1) \\ 
SFLM($J=0.4$, $L=0.60$, $M=0.4$) &- &0.600(3) &0.4000(1) &- &0.401(3) \\ 
SFLM($J=0.4$, $L=0.60$, $M=0.6$) &- &0.6003(4) &0.5999(1) &- &0.5995(4) \\ 
SFLM($J=0.4$, $L=0.60$, $M=0.7$) &- &0.6002(4) &0.6999(1) &- &.6992(9) \\ 
SFLM($J=0.6$, $L=0.60$, $M=0.3$) &0.5990(6) &0.602(3) &0.2998(2) &0.500(4) &0.499(1) \\ 
SFLM($J=0.6$, $L=0.60$, $M=0.4$) &0.6000(2) &0.600(2) &0.4001(1) &0.600(2) &0.600(1) \\ 
SFLM($J=0.6$, $L=0.60$, $M=0.6$) &0.5996(4) &0.6003(4) &0.5998(1) &0.7998(7) &0.798(1) \\ 
SFLM($J=0.6$, $L=0.60$, $M=0.7$) &0.6000(3) &0.5993(8) &0.7001(1) &0.900(4) &0.898(1) \\ 
VDP($H=0.3$) &0.4999(5) &0.498(4) &0.302(4) &0.300(5) &0.2996(7) \\ 
VDP($H=0.4$) &0.4999(5) &0.500(3) &0.400(3) &0.400(4) &0.4003(5) \\ 
VDP($H=0.6$) &0.4994(4) &0.500(2) &0.600(1) &0.599(2) &0.6008(4) \\ 
VDP($H=0.7$) &0.4993(4) &0.499(5) &0.701(4) &0.699(5) &0.6999(7) \\ 
\end{tabular}
\end{ruledtabular}
\end{table*}

\section{Application to financial markets}
\subsection{Financial Markets Data}
The data used for the analyses were one-minute valuations for the most actively traded exchange-traded funds (ETFs) in the US market extracted from PiTrading.com. We restrict consideration to the most recent 2500 trading days ($\sim$ 10 years). Intra-day trading is \emph{assumed} to be a realization of the same stochastic process. The data provide open, close, high, and low prices within every minute; we use close price for our analysis. Missing data, perhaps due to technical problems or errors, are replaced by the last recorded price. We studied the three most traded ETFs, the Dow Jones Industrial Average (DIA), S$\&$P 500 (SPY), and PowerShare NASDAQ-100 (QQQ). 

Stochastic processes underlying financial time series are represented using the \emph{return}
\begin{equation}
    X_t = \ln \frac{P_t}{P_0} \nonumber
\end{equation}
where $P_t$ is the price of a financial asset at time $t$,  and $P_0$ is a reference price, typically the price at the start of a session. The one-minute increments are
\begin{equation}
    \delta_{t} = X_t-X_{t-1} = \ln \frac{P_t}{P_{t-1}}. \nonumber
\end{equation}
Note that $X_{t=0} = 0$, which is the necessary condition for $X_t$ to scale, i.e. $X_t \stackrel{d}{=} X_1 $.

\subsection{Scaling regions}
\begin{figure}[b]
\includegraphics[width=0.45\textwidth]{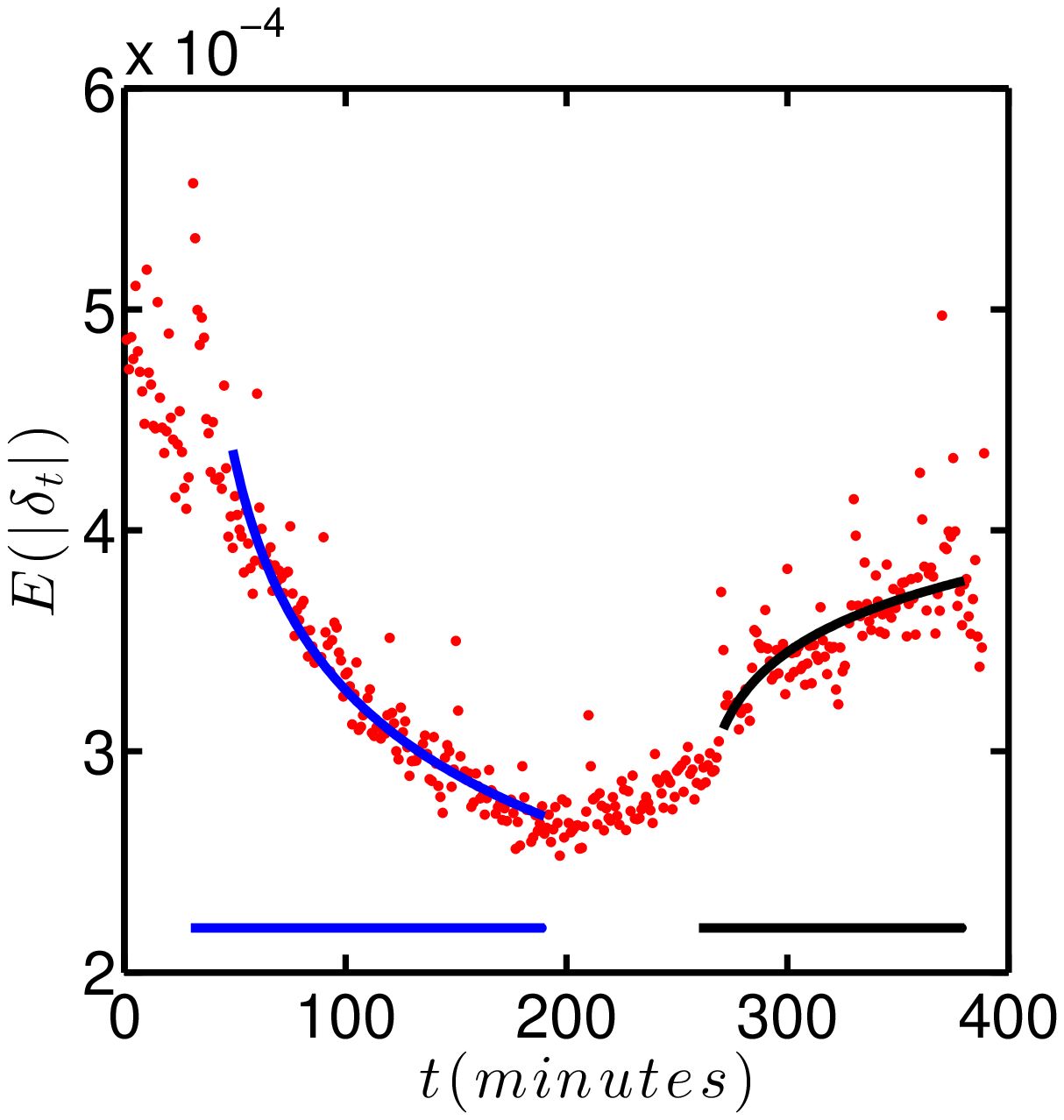}% Here is how to import EPS art
\caption{(Color online) Mean absolute value $E_{e}(|\delta_{t}|)$ of the increments of the SPY as a function of the time of day (beginning at 9:30 EST). Two scaling intervals where $E_{e}(|\delta_{t}|)$ can be fitted by power law $E_{e}(|\delta_{t}|) \sim t^{M-0.5}$, where $t$ is measured from the start of each interval.} 
\label{fig:SPYSeason}
\end{figure}

\emph{Intraday seasonality.} The analysis is predicated on the \emph{assumption} that intra-day variations of the return  follow a same stochastic process each day.  Consequently, return data from the 2500 trading days constitute the ensemble. In order to eliminate any drift within the trading day, the data is ``de-trended" by subtracting ensemble average $E_{e}(\delta_{t})$ at each $t$. The intraday  pattern of $E_{e}(|\delta_{t}|)$ shows that the stochastic behavior is non-stationary within the day. $E_{e}(|\delta_{t}|)$ of SPY (as well as for the other ETFs) appears to scale as a power law within two intervals during the day, the first following the opening of the market and and the second in the afternoon, see Figure~\ref{fig:SPYSeason}. The horizontal bars indicate the start and end of the two scaling intervals. The first ranges from 30 to 190 minutes and the second from 260 to 380 minutes from the start of the trading day.

\subsection{Estimation of the exponents}
\begin{figure*}
\includegraphics[width=0.90\textwidth]{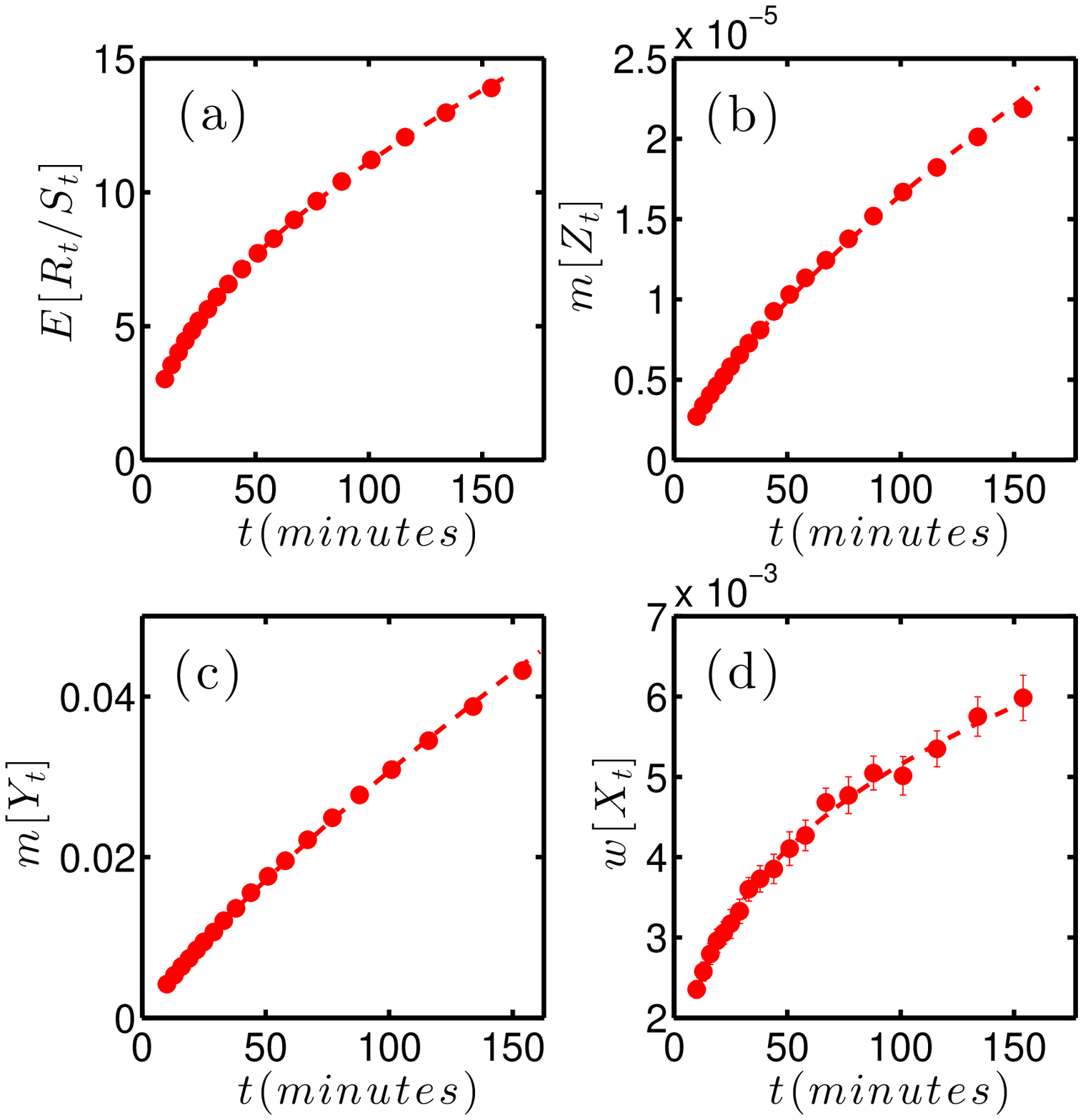}
\caption{(Color online) Nonlinear fit for the S\&P500 index SPY(30:190) using model $y(t) = a t^{\Omega^{\prime}} + b t^{\Omega^{\prime}-c}$ for finite-time corrections. (a) $y(t) = E_{e}[R_t/S_t]$, $\Omega^{\prime} = J = 0.500(2)$. (b) $y(t) = m[Z_t]$, $\Omega^{\prime} = 2L+2M-1 =0.587(4) $. (c) $y(t)=m[Y_t]$, $\Omega^{\prime} = M+0.5 = 0.790(2)$. (d) $y(t)=w[X_t]$, $\Omega^{\prime} = H = 0.298(7)$. The error bars in (a), (b), (c) are too small to be observed.}
\label{fig:SPYCombo}
\end{figure*}

The duration of the two scaling intervals are 180 and 120 minutes respectively. The lower cutoff $t_{min}$ for the finite-time analysis [Section 4(b)] was set to 10 minutes. The total number of points used for the analysis, with intervals given by Eqn.~\eqref{times}, is $t_{\#} = 60$. 
Methods outlined in the previous section were used for the analysis and Figure~\ref{fig:SPYCombo} shows the nonlinear fit for SPY(30:190). The indices extracted from the analyses are given in Table~\ref{tab:table3}, and conclusions include (1) $L \approx \frac{1}{2}$, implying that increments of the prices of ETFs are not from fat-tailed distributions, and (2) $J \approx \frac{1}{2}$ implying the absence of long-term memory. The latter is validated using the auto-correlation function which vanishes for time delays larger than 1 minute.  Furthermore, the relation $H = J+L+M-1$ is validated.
 
\begin{table*}
\caption{\label{tab:table3} Estimates for the exponents for exchange traded funds using the form $y(t) = a t^{\Omega^{\prime}} + b t^{\Omega^{\prime}-c}$ for finite-time corrections. SPY, DIA, and QQQ are abbreviations for the Standard and Poor 500 Index, the Dow Jones Industrial Average, and the PowerShare NASDAQ-100 Index respectively.}
\begin{ruledtabular}
\begin{tabular}{|c|c|c|c|c|c|}
Processes  &$J$ &$L$  &$M$ &$J+L+M-1$ &$H$ \\ \hline
SPY(30:190)   &0.500(2) &0.503(4) &0.290(2) &0.294(4) &0.298(7) \\
DIA(30:190)   &0.502(2) &0.497(6) &0.287(4) &0.286(4) &0.29(1) \\ 
QQQ(30:190)   &0.500(2) &0.501(6) &0.295(2) &0.296(5) &0.29(1) \\ 
SPY(260:380)   &0.500(2) &0.511(9) &0.552(4) &0.564(7) &0.55(1) \\ 
DIA(260:380)   &0.499(1) &0.50(1) &0.552(2) &0.56(1) &0.55(1) \\ 
QQQ(260:380)   &0.501(1) &0.50(1) &0.550(4) &0.56(1) &0.55(1) \\ 
\end{tabular}
\end{ruledtabular}
\end{table*}

\begin{figure*}
\includegraphics[width=0.9\textwidth]{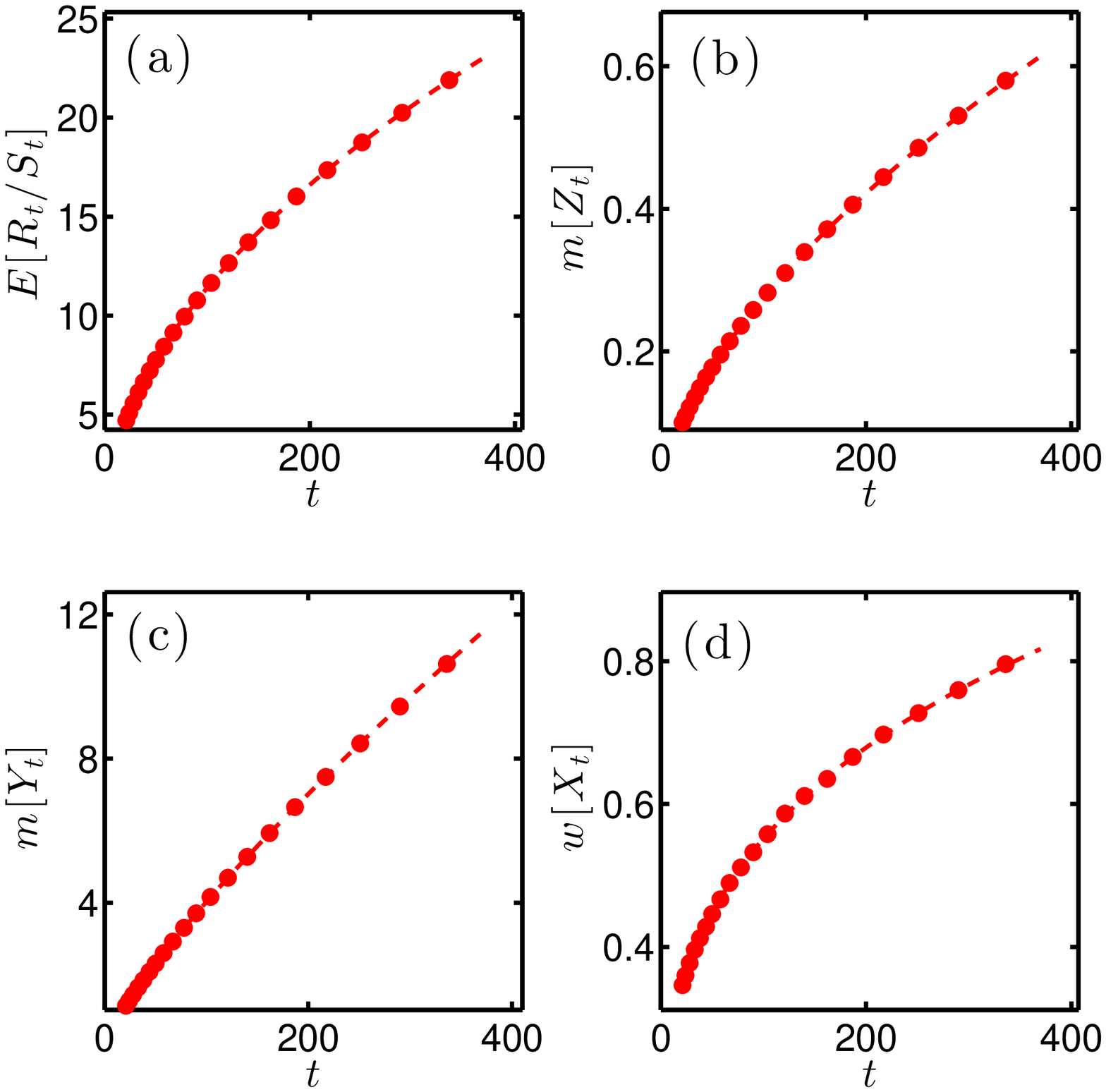}
\caption{(Color online) Nonlinear fit for a variable diffusion process (H=0.3) of length 370 using finite-time corrections of the form $y(t) = a t^{\Omega^{\prime}} + b t^{\Omega^{\prime}-c}$. (a) $y(t) = E_{e}[R_t/S_t]$, $\Omega^{\prime} = J = 0.499(3)$. (b) $y(t) = m[Z_t]$, $\Omega^{\prime} = 2L+2M-1 = 0.596(4)$. (c) $y(t)=m[Z_t]$, $\Omega^{\prime} = M+0.5 = 0.8002(2)$. (d) $y(t)=w[Y_t]$, $\Omega^{\prime} = H = 0.301(2)$. The errors bars are too small to be visible.}
\label{fig:VDPCombo}
\end{figure*}

The analyses outlined in Section IV were carried out for 100,000 ensembles of length 1 million. In contrast, financial market analysis was conducted for an ensemble of size 2500, and the intervals were 180 and 120 minutes respectively. One may inquire if the information extracted from these short processes through finite-size corrections are reliable. In order to address this issue, we re-computed the indices for variable diffusion process over a time interval $t_{max}$. As described below, $t_{max}$ is evaluated using the relaxation of the finite-time indices. Specifically, we write Eqn.~\eqref{FTC0} as 
\begin{equation}
E_{e}[R/S(t)] = a t^{J}\left[1 - \left(\frac{t}{\tau}\right)^{-c} \right]   \nonumber
\end{equation}
where $\tau = (-\frac{b}{a})^{\frac{1}{c}} = 0.18(1)$ is a time-scale for convergence of the index. In order to estimate $t_{max}$ for the VDP, we note that the nonlinear fit for $J$ for SPY(30:190) gives $\tau = 0.42(5)$. The corresponding analysis of the VDP ($H=0.3$) should be of length $t_{max} = \frac{0.42}{0.18}* 160 \approx 370$, $t_{min} = \frac{0.42}{0.18}*10 \approx 20$. Figure~\ref{fig:VDPCombo} shows the nonlinear fit for this VDP with $t_{max} = 370$. The estimated exponents are $J =0.499(3)$, $L=0.497(6)$, $M = 0.300(4)$, $J+L+M-1 = 0.297(5)$, $H= 301(2)$. We thus infer that exponents computed for financial markets are reliable. 

\section{Conclusions}
There are increasing numbers of examples, ranging from variations in biological systems~\cite{Iannaccone1995} and thermal fluctuations in turbulence~\cite{castaing1989scaling} to price variations in financial markets~\cite{mantegna1995scaling}, where probability distributions associated with a stochastic process exhibit anomalous scaling and are non-Gaussian. The anomalies can have different origins, and the goal of the work reported here is to disentangle them. 
Previous studies on stationary processes~\cite{mandelbrot2002gaussian,Eliazar2013} had established that infinite variance of increments (Noah effect) and long-time correlations between increments  (Joseph effect) are two sources of the anomaly. However, these studies failed to recognize that time-dependence of the increments themselves can also be a source of anomalous scaling. 
In this paper, we showed how scaling of the increments with time, referred to as the Moses effect, can also contribute to anomalous scaling. 
Noah, Joseph, and Moses effects, characterized by $L$, $J$ and $M$ respectively, are independent and the overall scaling of probability distributions, quantified by the Hurst exponent $H$, is given by $H = L+J+M-1$.  As was emphasized, definitions of the exponents require the use of an ensemble of (nominally identical) stochastic trajectories when the underlying processes are time dependent. 

Numerical approaches of time series analysis
to accurately estimate each of the four scaling exponents independently were introduced. They are based on the use of medians and quantiles, which is especially appropriate for probability distributions lacking finite variance.
These methods account for finite-time power-law corrections to scaling. The fact that the four indices can
be measured independently allows the scaling relation that connects them to be verified, providing a stringent
numerical check on the accuracy of the time-series analysis. The new numerical techniques were applied to a variety of 
different stochastic processes, including ones with both stationary and non-stationary increments, with and 
without long-time auto-correlations, and with both finite and infinite increment variance, 
to demonstrate the role of each effect toward anomalous scaling.

Financial time series of 
exchange-traded funds (ETFs) were analyzed as an application of the methods introduced here. As has been found to be the case for other financial markets~\cite{bassler2007nonstationary,Seemann2012,Hua2015},
the intraday prices of ETFs can be considered to be governed by non-stationary stochastic processes
that repeat each trading day.     
We find two intervals where the underlying stochastic process scales anomalously with $H \neq 1/2$, which
is often associated with a violation of the efficient market hypothesis (EMH).  
However, we find that $L = 1/2$ and $J = 1/2$; \emph{i.e.}, neither Noah nor  
Joseph effects are observed in financial markets. 
The deviation from $H=1/2$ results solely from the Moses effect ($M \neq 1/2$).
Previously, it was proposed that the true test of the EMH should be the lack of correlations, {i.e.},  $J = 1/2$, 
and not $H=1/2$~\cite{EMH1}.
Therefore, our analysis reiterates that ETF markets satisfy the EMH, 
despite the fact that they exhibit anomalous scaling.
Finally, consistent with other recent studies of intraday trades 
in financial markets~\cite{bassler2007nonstationary, mccauley2008martingales1, mccauley2008martingales2}, we found that a variable diffusion process accurately models 
the scaling behavior in the two scaling intervals.

Although the scaling exponents defined here characterize the sources of anomalous scaling of the distribution functions, they do not uniquely identify higher order statistic or the underlying stochastic processes themselves. As an example, a scaled Brownian process and a variable diffusion process can have the same exponents as the first stage of stock markets (\emph{i.e.,} with $J=\half$, $H=\half$, and $M=0.3$). As discussed in the Supplemental Material at [URL will be inserted by publisher], VDP exhibits volatility clustering (\emph{i.e.,} the absolute values and the squares of increments exhibit long-time correlations) while the corresponding scaled BM does fails to do so. Volatility clustering is one of the well-known \emph{stylized facts} on financial market dynamics~\cite{VolClustering}.

It would be interesting to apply the methods of time series analysis developed 
here to other, more physical, recurring stochastic processes with 
non-stationary increments.
For example, the amount of daily precipitation recorded at a fixed location~\cite{kantelhardt2006long, kantelhardt2003multifractality, brunsell2010multiscale, rehman2009study, richardson1981stochastic}
may be amenable to 
such analysis. If underlying (stochastic) process is assumed to repeat itself each year, an ensemble can be constructed using the data for each year.  
Similarly, the daily, or monthly abundance of solar flares~\cite{lepreti2000persistence, oliver1998there, rehman2009study, richardson1981stochastic} may also be
amenable to our methods of analysis. Solar activity is known to have an eleven year cycle, 
so that the days or months at the same phase each cycle may form an ensemble.
The approach may also be useful for analyzing hard turbulence~\cite{castaing1989scaling}; 
here the temperature variation at a given location may be taken to be a 
stochastic process with non-stationary
increments~\cite{rehman2009study, richardson1981stochastic}. 
The process may then be considered to repeat after a non-periodic
triggering event, such as a boundary layer separation~\cite{castaing1989scaling}. In this case, the temperatures
at a given time following the triggering event form an ensemble.
In each of these systems, it would be interesting to learn if the scaling is anomalous and,
if so, which of the Noah, Joseph, and Moses effects, or what combination thereof, 
leads to the anomaly.
\\

This work was supported by NSF-DMR-1507371 (KB) and NSF-IOS-1546858 (KB).

\nocite{*}

\bibliography{reference}% Produces the bibliography via BibTeX.
\end{document}